\documentclass[a4paper,twoside]{article}
\usepackage[latin1]{inputenc} 
\usepackage[T1]{fontenc} 
\usepackage{RR}
\usepackage{hyperref}

\usepackage{paralist}
\usepackage{graphicx}
\usepackage{subfigure}
\usepackage{listings}
\usepackage{amsmath}
\usepackage{wasysym}
\usepackage{amssymb}
\usepackage{amsfonts}
\usepackage{moreverb}
\usepackage{url}
\usepackage{cite}
\usepackage{multirow}
\usepackage{hyperref}
\usepackage[all,2cell,ps]{xy}
\usepackage{amsthm}
\theoremstyle{plain}
\newtheorem{theorem}{Theorem}
\newtheorem{definition}{Definition}
\newtheorem{property}{Property}
\theoremstyle{definition}
\newtheorem{example}{Example}

\usepackage[T1]{fontenc}
\usepackage{alltt}
\usepackage{verbatim}
\usepackage{moreverb}
\usepackage{amssymb}
\usepackage{listings}
\usepackage{float}
\usepackage[linesnumbered,vlined]{algorithm2e}
\usepackage{algorithmic}
\usepackage{graphicx}
\usepackage{wrapfig}
\usepackage{textcomp}
\newenvironment{algo}{%
\algorithm
}{%
\endalgorithm
}

\RRNo{7870}
\RRdate{January 2012}
\RRauthor{
Houari Mahfoud \thanks{University of Nancy 2 \&  INRIA Nancy Grand Est 
(\texttt{Houari.Mahfoud@inria.fr}).} and
Abdessamad Imine\thanks{University of Nancy 2 \&  INRIA Nancy Grand Est
(\texttt{Abdessamad.Imine@inria.fr}).}
}
\authorhead{H. Mahfoud and A. Imine}
\RRtitle{Une Approche G\'en\'erale pour S\'ecuriser l'Acc\`es et la Mise \`a jour des Donn\'ees XML}
\RRetitle{A General Approach for Securely Querying and Updating XML Data} 
\titlehead{A General Approach for Securely Querying and Updating XML Data}
\RRresume{Durant ces derni\`eres ann\'ees, plusieurs travaux ont propos\'e des mod\`eles de contr\^ole d'acc\`es pour s\'ecuriser l'acc\`es en lecture aux donn\'ees XML, bas\'es seulement 
sur des DTDs non-r\'ecursives. Le contr\^ole d'acc\`es XML consid\'erant les op\'erations de mise \`a jour n'a pas re\c cu suffisamment d'attention.
Dans ce papier, nous pr\'esentons un mod\`ele g\'en\'eral pour sp\'ecifier le contr\^ole d'acc\`es aux donn\'ees XML moyennant des primitives de mise \`a jour du \emph{W3C XQuery 
Update Facility}. Notre approche pour enforcer ces sp\'ecifications de mise \`a jour est bas\'ee sur la notion de r\'e\'ecriture des requ\^etes (\emph{query rewriting} en anglais) o\`u chaque 
op\'eration de mise \`a jour, d\'efinie par rapport \`a une DTD arbitraire (r\'ecursive ou non), est r\'e\'ecrite en une autre op\'eration s\^ure afin qu'elle soit \'evalu\'ee seulement sur des donn\'ees 
XML modifiables par l'utilisateur qui a soumis l'op\'eration. 
Nous \'etudions dans la deuxi\`eme partie de ce rapport la s\'ecurisation des op\'erations de mise \`a jour XML en pr\'esence des droits de lecture sp\'ecifi\'es sous forme d'une \emph{vue de s\'ecurit\'e}. 
Pour un document XML, une vue de s\'ecurit\'e permet de rep\'esenter pour chaque classe d'utilisateurs les parties du document dont ils sont autoris\'es \`a voir. 
Nous montrons qu'une op\'eration de mise \`a jour d\'efinie par rapport \`a une vue de s\'ecurit\'e peut entra\^iner des divulgations des donn\'ees 
confidentielles cach\'ees par cette vue, si elle n'est pas soigneusement r\'e\'ecrite en tenant compte des droits de lecture et de mise \`a jour. 
Pour pallier \`a ce probl\`eme, nous d\'ecrivons une solution qui permet de pr\'eserver la confidentialit\'e et l'int\'egrit\'e des donn\'ees XML.
}
\RRabstract{
Over the past years several works have proposed access control models for XML data where only read-access rights over 
non-recursive DTDs are considered. A few amount of works have studied the access rights for updates.
In this paper, we present a general model for specifying access control on XML data in the presence of update operations of W3C XQuery Update Facility. 
Our approach for enforcing such updates specifications is based on the notion of \emph{query rewriting} where 
each update operation defined over arbitrary DTD (recursive or not) is rewritten to a safe one in order to be evaluated only 
over XML data which can be updated by the user.
We investigate in the second part of this report the secure of XML updating in the presence of read-access rights specified by a \emph{security views}. 
For an XML document, a security view represents for each class of users all and only the parts of the document these users 
are able to see. We show that an update operation defined over a security view can cause disclosure of sensitive data hidden by this view if it is not thoroughly rewritten 
with respect to both read and update access rights. 
Finally, we propose a security view based approach for securely updating XML in order to preserve the confidentiality and integrity of XML data.
}
\RRmotcle{Cont\^ole d'acc\`es XML, Vues de s\'ecurit\'e XML, Mise \`a jour XML, R\'e\'ecriture des requ\^etes, XPath, XQuery, Confidentialit\'e et Int\'egrit\'e}
\RRkeyword{XML access control, XML security views, XML updating, Query rewriting, XPath, XQuery, Confidentiality and Integrity}
\RRprojets{CASSIS}
\RCNancy 

\begin{document}
\makeRR   
\tableofcontents
\newpage
\section{Motivation}\label{Sect1}
The XQuery Update Facility language \cite{ref16} is a recommendation of W3C that provides facility to modify some parts of an XML document and leaving 
the rest unchanged, and this through different update operations, e.g., insert, replace, or delete some nodes of a given XML document.
The security requirement is the main problem when manipulating XML documents.
An XML document may be queried and/or updated simultaneously by different users. For each class of users some rules can be defined to specify parts of the 
document which are accessible to the users and/or updatable by them. 
A bulk of work has been published in the last decade to secure the XML content, but only 
read-access rights has been considered over non-recursive DTDs \cite{ref4}, \cite{ref6}, \cite{ref19}. 
Moreover, a few amount of works have considered update rights.  

In this paper, we investigate a general approach for securing XML update operations of the XQuery Update Facility language. 
Abstractly, for any update operation posed over an XML document, we ensure that the operation is performed only on XML nodes updatable by the user 
and no sensitive information can be deduced via this operation.
Addressing such concerns requires first a specification model to define update constraints 
and a flexible mechanism to enforce these constraints at update time.

\begin{figure}[t!]
\centering
\includegraphics[width=8cm]{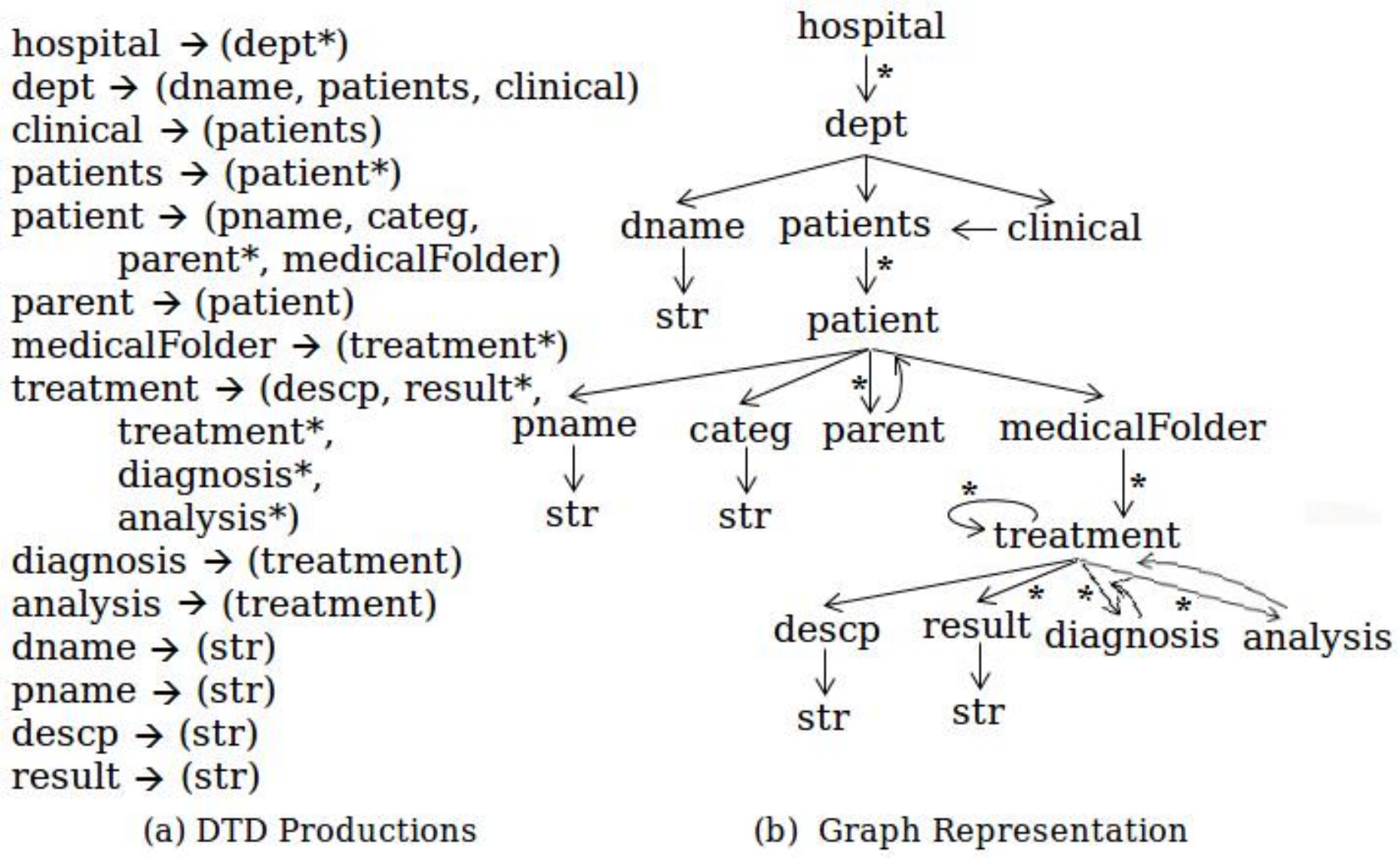}
\caption{Hospital DTD.}
\label{hospitalDTD}
\end{figure}

\begin{figure}[t!]
\centering
\includegraphics[width=9.4cm]{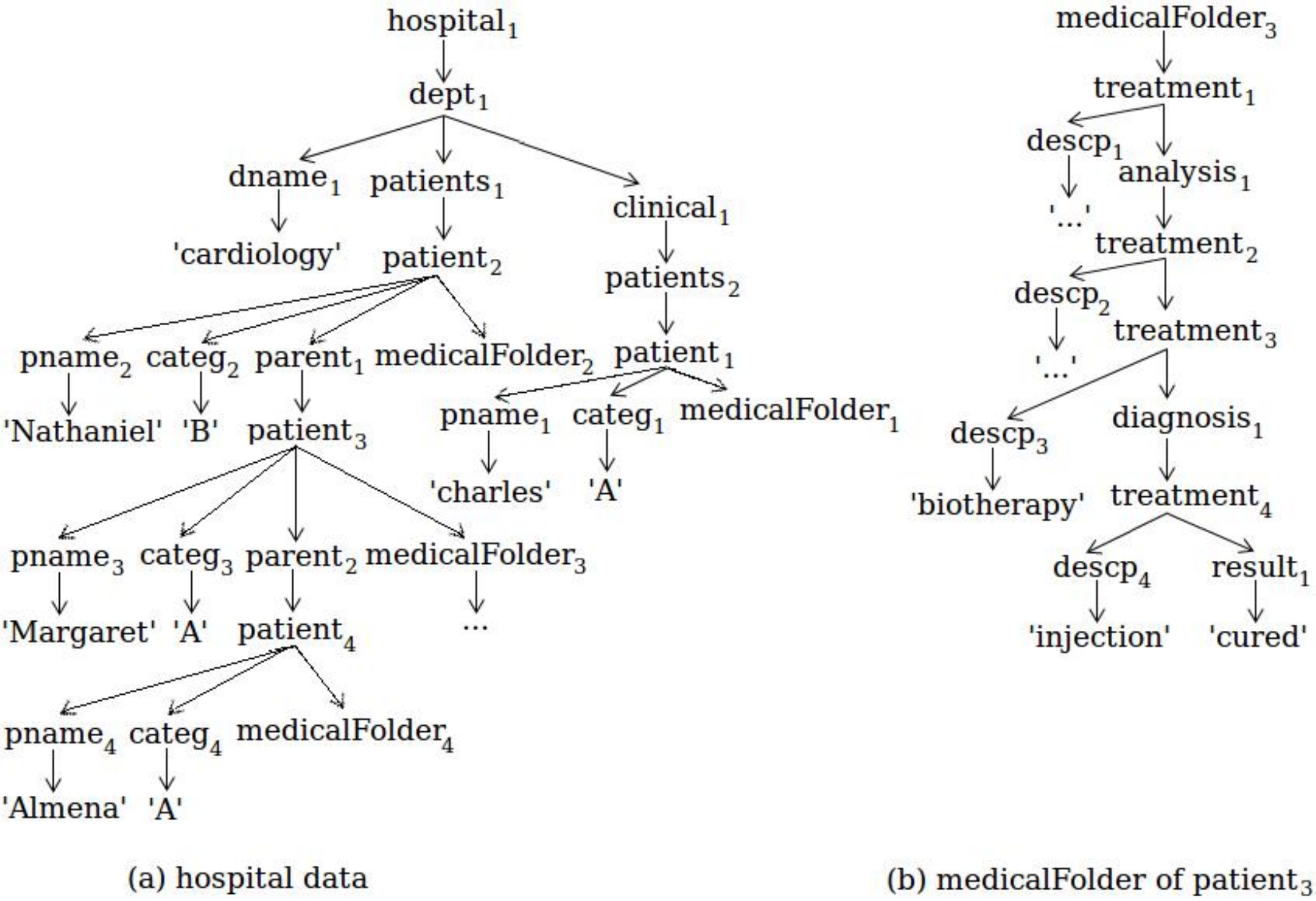}
\caption{Example of XML Document.}
\label{hospitalDATA}
\end{figure}

We now present our motivating example for controlling update access. 
Consider the recursive DTD\footnote{A DTD is recursive iff at least one of its elements is defined (directly or indirectly) in terms of itself.} of a hospital 
depicted as a graph in Fig. \ref{hospitalDTD}(b) (we refer to this DTD throughout the paper to illustrate our examples).
An XML document conforming to this DTD consists of different departments (\emph{dept}) defined by a name \emph{dname} and each department includes 
patients of the hospital and other patients coming from some clinics (patients under \emph{clinical} element). For each patient (with name \emph{pname} and 
category \emph{categ}), the hospital maintains a medical history of its parents (\emph{parent}) and a medical folder (\emph{medicalFolder}) which includes 
all treatments done for this patient (\emph{treatment} can be \emph{analysis} or \emph{diagnosis}); \emph{descp} and \emph{result} represent the 
description and the result of the treatment respectively. 
The treatment data is organized into two groups depending on whether the treatment has been done in some laboratories 
(\emph{analysis} treatments) or not (the \emph{diagnosis} treatments). Each \emph{dname}, \emph{pname}, \emph{categ}, \emph{descp}, and \emph{result} has a 
single text node (\texttt{PCDATA}) as its child.
An instance of the hospital DTD is given in Fig. \ref{hospitalDATA}. Due to space limitation, this instance is split into 
Figures \ref{hospitalDATA} (a) and (b), where Fig. \ref{hospitalDATA}(b) represents 
the medical folder of $patient_{3}$. 

Suppose that the hospital wants to impose an update policy that allows the doctors to update all treatments data (e.g., add some treatment results) except those of analysis 
(done outside the hospital). According to this policy, only the nodes $treatment_{1}$ and $treatment_{4}$ of Fig. \ref{hospitalDATA}(b) can be updated. As the nodes 
$treatment_{2}$ and $treatment_{3}$ are analysis treatments they cannot be updated.\medskip

\noindent \textit{\textbf{Problem 1.}} The existing access control approaches are unable to specify the above policy.
The model given in \cite{ref19} consists in annotating the schema of the document by 
different update constraints, like putting attribute \texttt{@insert=$Y$} in element type $treatment$ of the hospital DTD to specify 
that some data can be inserted into nodes of type $treatment$. However, only local annotations (the update concerns only the node and not its descendants) are used which is not sufficient to define some update policies. 
For instance, to enforce the hospital update policy imposed, the analysis treatment data (i.e., nodes $treatment_2$ and $treatment_3$) cannot 
be discarded from doctors's updates by the model introduced in \cite{ref19} even by using XPath upward-axes. Specifically, the annotation 
\texttt{@insert=[not(ancestor::$analysis$)]} over element type $treatment$ is not the adequate constraint since it makes node $treatment_4$ not updatable.

In the XACU$^{annot}$ language presented in \cite{ref3}, an update annotation over an element type of the DTD is defined with 
a full path from the DTD root to this element. 
E.g., the annotation $ann(hospital/patients/patient,\texttt{insert})$=$Y$ specifies that some nodes can be inserted under hospital patients. 
However, the XACU$^{annot}$ language cannot be applied in the presence of recursive DTDs. 
For instance, due to recursion, the hospital update policy given above cannot be defined since the paths denoting updatable $treatment$ nodes 
(not done during $analysis$) stand for an infinite set of paths.
As we will see in the next, this set of paths can be expressed using the $Kleene$ star operator (\texttt{*}) which 
cannot be expressed in the standard XPath as outlined in~\cite{ref23,ref24}.
To our knowledge, no model exists for specifying update policies over recursive DTDs.\medskip

\noindent \textit{\textbf{Problem 2.}} 
For each update operation, an XPath expression is defined to specify the XML data at which the update is applied. 
To enforce rights restriction imposed by an update policy, 
the \emph{query rewriting} principle can be applied where each 
update operation (i.e., its XPath expression) is rewritten according to the update rights into a safe one in order to be evaluated only over parts of the XML data updatable 
by the user. However, this rewriting step is already challenging for a small class of XPath. 
Consider the downward fragment of XPath which supports $child$ and $descendant$ axes, union and complex predicates. We show that, in case 
of recursive DTDs, an update operation defined in this fragment cannot be rewritten safely. 
More specifically, a safe rewriting of the XPath expression of an update operation can stand for an infinite set of paths 
which cannot be expressed in the downward fragment of XPath. 
To overcome this rewriting limitation, some solutions have been proposed \cite{ref5, ref8} based on the Regular XPath to express 
safe recursive paths. However, these solutions remain a theoretical achievement since no tool exists to 
evaluate Regular XPath expressions. Thus, no practical solution exists for enforcing update policies in the presence of recursive DTDs.\medskip

\noindent \textit{\textbf{Problem 3.}} We discuss finally the interaction between read and update privileges. 
For each class of users, some read-access rights can be defined to prevent 
access to sensitive data of the XML document. Moreover, update rights can be imposed to specify parts of the document which can be updated by these users. 
In this case, we show that rewriting an update operation by considering simply the update rights is not sufficient to 
make XML updates secure. In other words, an update operation can be safe w.r.t update policy; but, evaluating 
this operation over the XML document can make disclosure of sensitive data.
For instance, suppose that the doctors can update all data in the hospital, but they can see only patients of category ``$A$''. 
According to this read-access right, a view can be computed from the instance of Fig. \ref{hospitalDATA}(a) by hiding node $patient_2$ and its children 
nodes ($pname_2$, $categ_2$, $parent_1$, and $medicalFolder_2$). Thus, node $patient_3$ is shown to the doctors since its category is ``$A$'' and it 
appears as an immediate child of node $patients_1$. 
Consider now the update \textbf{delete} $descendant$::$patients$[$patient$[$pname$='$Nathaniel$']]/$descendant$::$result$ that consists in removing all $result$ nodes
provided that patient $Nathaniel$ exists. This update is safe w.r.t the update policy defined 
above. However, if the execution of this update succeeds 
then the user can deduce that patient $Nathaniel$ is currently residing in the hospital and his medical data 
is confidential.
Consequently, the interaction between read and update privileges should be thoroughly designed in order to preserve confidentiality and integrity properties.\medskip

We present in the following our main contributions of this work proposed to deal with the previous problems.\medskip

\noindent \textbf{Our Contributions.} Our first contribution is an expressive model for specifying XML update policies, based on the primitives of XQuery 
Update Facility, and over arbitrary DTDs (recursive or not). Given a DTD $D$, 
we annotate element types of $D$ with different update rights to specify restrictions on updating XML documents conform to $D$ through some update 
operations (e.g., deny insertion of new nodes of type $analysis$ under $treatment$ nodes).
We propose a new model that supports inheritance and overriding of update privileges and overcomes expressivity limitations of existing 
models (see \textbf{Problem 1}). Our approach for enforcing such update policies is based on the notion of \emph{query rewriting}. 
However, to overcome the rewriting limitation presented above as \textbf{Problem 2}, we 
investigate the extension of the downward fragment of XPath by some axes and operators. Based on this extension, our second contribution is an algorithm 
that rewrites any update operation defined in the downward fragment of XPath into another one defined in the extended fragment to 
be safely evaluated over the XML data. 
We discuss in the second part of this paper our solution to deal with \textbf{Problem 3}. 
We propose a general approach to secure update operations defined over a (recursive) security view without disclosure of sensitive data hidden by this view (i.e., to preserve confidentiality and integrity 
of the XML data, each update operation over the view must be rewritten to be safe w.r.t both read and update rights). 
To our knowledge, this paper presents the first model for specifying and enforcing update policies using the XQuery update operations and in the presence of arbitrary DTDs (resp. arbitrary security views).\medskip

\noindent \textbf{Related Work.}
During the last years, several works have proposed access control models to secure XML content, but only 
read-access has been considered over non-recursive DTDs \cite{ref4,ref6,ref19}. There has been a few amount of works on securing XML data by 
considering the update rights. 
Damiani et al. \cite{ref19} propose an XML access control model for update operations of the XUpdate language. They annotate the XML schema 
with the read and update privileges, and then the annotated schema is translated into two automatons defining read and update policies 
respectively, which are used to rewrite any access query (resp. update operation) over the XML document to be safe. 
However, the update policy is expressed only with local annotations which is not sufficient to specify 
some update rights (see \textbf{Problem 1}). Additionally, the automaton processing cannot be successful when rewriting access 
queries (resp. update operations) defined over recursive schema (i.e., recursive DTD).

Authors of \cite{ref3} propose an XML update access control model based on the XQuery update operations. 
A set of XPath-based rules is used to specify, for each update operation, the XML nodes that can be updated by the user using this operation.
These rules are translated into annotations over element types of the DTD (if exists) to present an annotation-based model called XACU$^{annot}$. 
However this translation is possible only in case of non-recursive DTDs.

Consider the read-access control models. Unlike the secure of XML querying over non-recursive security views, 
the problem posed by the recursion (i.e., XPath query rewriting is not always possible under recursive 
security views) has not received a more attention. 
To overcome this problem, some authors \cite{ref5,ref8} propose rewriting approaches based on the non-standard language, ``Regular XPath'', 
which is more expressive than XPath and makes rewriting possible under recursion. 
However, no practical system exists of both proposed approaches\footnote{According to \cite{ref11} the SMOQE system proposed in \cite{ref13} has been 
removed because of conduction of future researches.}, and in general, no tool exists to evaluate Regular XPath queries over XML data. 
Thus, the need of a rewriting system of XPath queries (resp. update operations) over 
recursive security views remains an open issue.\medskip

\noindent \textbf{Outline.} The remainder of the paper is organized as follows. Section \ref{Sect2} presents basic notions on DTD, XPath, and XML update 
operations considered in this paper. We describe in Section \ref{Sect3} our specification model of update. 
Our approach for securing update operations is detailed in Section \ref{Sect4}. 
We recall the notion of security view in Section \ref{Sect5} and present a view-based approach to secure updating of XML 
documents over (recursive) security views. Finally, we conclude this paper in Section \ref{Sect6}.

\section{Preliminaries}\label{Sect2}
This section briefly reviews some basic notions tackled throughout the paper.\medskip

\noindent\textbf{DTDs.} Without loss of generality, we represent a DTD $D$ by (\emph{Ele}, \emph{Rg}, \emph{root}), where \emph{Ele} is a finite set 
of \emph{element types}; \emph{root} is a distinguished type in \emph{Ele} called the \emph{root type}; 
\emph{Rg} is a function defining element types such that for any \emph{A} in \emph{Ele}, \emph{Rg}($A$) is a regular expression $\alpha$ defined as follows:

\begin{center}
$\alpha$ \texttt{:= str} | $\epsilon$ | $B$ | $\alpha$','$\alpha$ | $\alpha$'|'$\alpha$ | $\alpha$\texttt{*}
\end{center}

\noindent where \texttt{str} denotes the text type \texttt{PCDATA}, $\epsilon$ is the empty word, \emph{B} is an element type in \emph{Ele}, and finally 
$\alpha$','$\alpha$, $\alpha$'|'$\alpha$, and $\alpha$\texttt{*} denote concatenation, disjunction, and the Kleene closure respectively. 
We refer to \emph{A $\rightarrow$ Rg}($A$) as the \emph{production} of \emph{A}. For each element type \emph{B} occurring in \emph{Rg}($A$), we refer to 
\emph{B} as a \emph{subelement type} (or \emph{child type}) of \emph{A} and to \emph{A} as a \emph{superelement type} 
(or \emph{parent type}) of \emph{B}. 
A DTD $D$ is said \emph{recursive} if some element type \emph{A} is defined in terms of itself directly or indirectly.

We use graph representation to depict our DTDs. 
For instance, Fig. \ref{hospitalDTD} represents (a) the productions of the hospital DTD; and (b) its graph representation corresponding.\medskip

\noindent \textbf{XML Documents.} We model an XML document with an unranked ordered finite node-labeled tree, called \emph{XML Tree}. 
Let $\Sigma$ be a finite set of node labels, an XML tree $T$ over 
$\Sigma$ is a structure defined as\cite{ref23}: $T$=$(N,R_{\downarrow},R_{\rightarrow},L)$ where $N$ is the set of nodes, 
$R_{\downarrow}$ $\subseteq$ $N\times N$ is a child relation, 
$R_{\rightarrow}$ $\subseteq$ $N\times N$ is a successor relation on (ordered) siblings, and $L: N\rightarrow \Sigma$ is a function assigning 
to every node its label. 
$R_{\uparrow}$ and $R_{\leftarrow}$ denote the converse of the relations $R_{\downarrow}$ and $R_{\rightarrow}$ respectively. 
For instance, $R_{\leftarrow}$ $\subseteq$ $N\times N$ is a predecessor relation on (ordered) siblings.

An XML document $T$ conforms to a DTD $D$ if the following conditions hold: 
(\emph{i}) the root of $T$ is the unique node labeled with \emph{root}; 
(\emph{ii}) each node in $T$ is labeled either with an \emph{Ele} type $A$, called an \emph{A element}, or with \texttt{str}, called a \emph{text node}; 
(\emph{iii}) for each $A$ element with $k$ ordered children $n_1,...,n_k$, the word $L(n_1),...,L(n_k)$ belongs to the regular language defined by \emph{Rg}($A$); 
(\emph{iv}) each text node carries a string value (\texttt{PCDATA}) and is the leaf of the tree. 
We call $T$ an instance of $D$ if $T$ conforms to $D$.\medskip

\noindent\textbf{XPath Queries.} We consider a small class of XPath \cite{ref17} queries, referred to as $\mathcal{X}$ and defined as follows:
\begin{alltt}
 \textit{p} := \(\alpha\)::\textit{lab} \(|\) \textit{p}[\textit{q}] \(|\) \textit{p}/\textit{p} \(|\) \textit{p} \(\cup\) \textit{p}
 \textit{q} := \textit{p} \(|\) \textit{p}/\textit{text()}='\(c\)' \(|\) \textit{q} \(and\) \textit{q} \(|\) \textit{q} \(or\) \textit{q} \(|\) \(not\) (q)
 \textit{\(\alpha\)} := \(\varepsilon\) \(|\) \(\downarrow\) \(|\) \(\downarrow\sp{+}\) \(|\) \(\downarrow\sp{*}\)
\end{alltt}

\noindent where $p$ denotes an XPath query and it is the start of the production, 
\emph{lab} refers to element type or $*$ (that matches all types), $\cup$ stands for union, 
$c$ is a string constant, $\alpha$ is the XPath axis relations, and $\varepsilon$, $\downarrow$, $\downarrow^{+}$ and $\downarrow^{*}$ denote $self$, 
$child$, $descendant$ and $descendant$-$or$-$self$ axis respectively. 
Finally the expression $q$ enclosed in [.]~is called a \emph{qualifier} (\emph{predicate} or \emph{filter}).

Let $n$ be a node in an XML tree $T$. The evaluation of an XPath query $p$ at node $n$, called \emph{context node} $n$, results in a set of 
nodes which are reachable via $p$ from $n$, denoted by $n$\textlbrackdbl$p$\textrbrackdbl. A qualifier $q$ is said valid at 
context node $n$, denoted by $n\vDash q$, iff one of the following conditions holds: 
(\textit{i}) $q$ is an atomic predicate and $n$\textlbrackdbl$q$\textrbrackdbl$\;$ is nonempty, i.e., there exists a node reachable via $q$ from $n$; 
(\textit{ii}) $q$ is given by \emph{p/text}()='$c$' and $n$\textlbrackdbl$p$\textrbrackdbl$\;$ contains a node which has a child text node whose string 
value is $c$; (\textit{iii}) $q$ is a boolean expression and it is evaluated to true at $n$, e.g., predicate $not(q)$ is valid at $n$ 
iff $n$\textlbrackdbl$q$\textrbrackdbl$\;$ is empty.\medskip

Theoretically, this XPath fragment (called \emph{downward} fragment) has some interesting decision results \cite{NevenS06}. 
Practically, it is commonly used and is is essential to XQuery, XSLT and XML Schema \cite{ref5}. 
Authors of \cite{ref5} have shown that in case of recursive security views, 
the fragment $\mathcal{X}$ is not closed under query rewriting (i.e., some access queries defined in $\mathcal{X}$ cannot be rewritten to be safe). 
This problem is known as \emph{XPath query rewriting problem}. 
We show that the same problem is encountered in controlling update operations and we propose a solution based on the extension of 
fragment $\mathcal{X}$ as follows:

\begin{alltt}
 \textit{p} := \(\alpha\)::\textit{lab} \(|\) \textit{p}[\textit{q}] \(|\) \textit{p}/\textit{p} \(|\) \textit{p} \(\cup\) \textit{p} \(|\) \textit{p}[\textit{n}]
 \textit{q} := \textit{p} \(|\) \textit{p}/\textit{text()}='\(c\)' \(|\) \textit{q} \(and\) \textit{q} \(|\) \textit{q} \(or\) \textit{q}
      \(|\) \(not\) (q) \(|\) \textit{p}=\(\varepsilon\)::\textit{lab}
 \textit{\(\alpha\)} := \(\varepsilon\) \(|\) \(\downarrow\) \(|\) \(\downarrow\sp{+}\) \(|\) \(\downarrow\sp{*}\) \(|\) \(\uparrow\) \(|\) \(\uparrow\sp{+}\) \(|\) \(\uparrow\sp{*}\)
\end{alltt}

\noindent we enrich $\mathcal{X}$ by the upward-axes \emph{parent} ($\uparrow$), \emph{ancestor} ($\uparrow^{+}$), and 
\emph{ancestor}-\emph{or}-\emph{self} ($\uparrow^{*}$), the \emph{position} and the \emph{node comparison} predicates.
The position predicate, defined with \texttt{[$n$]}(\emph{n} $\in$ N), is used to return the $n^{th}$ node from an ordered set of nodes. 
For instance, since we model an XML document with an ordered tree, the query $\downarrow$::$*[1]$ over a node $n$ returns its first child node. 
The \emph{node comparison} predicate [$p_{1}$=$p_{2}$] is valid at a node $n$ only if the evaluation of the right and left XPath queries 
at $n$ result in exactly the same single node. For example, the predicate $[\uparrow$::$*$=$\uparrow^{+}$::$*[1]]$ is valid at any node $n$ 
having a parent node. 
We summarize this extension by the following subsets $\mathcal{X}^{\Uparrow}$ ($\mathcal{X}$ with upward-axes), 
$\mathcal{X}^{\Uparrow}_{[n]}$ ($\mathcal{X}^{\Uparrow}$ with position predicate), and $\mathcal{X}^{\Uparrow}_{[n,=]}$ 
($\mathcal{X}^{\Uparrow}_{[n]}$ with node comparison predicate). 


In our case, fragment $\mathcal{X}$ is used only to formulate update operations (resp. access queries) and to define our 
update policies (resp. access policies), 
while we will explain later how the augmented fragments of $\mathcal{X}$ defined above can be used to avoid the XPath query rewriting problem.\medskip

\noindent \textbf{XML Update Operations.}
We review some update operations of the W3C XQuery Update Facility recommendation \cite{ref16} (abbreviated as XUF). 
We study the use of the following operations: \emph{insert}, \emph{delete}, and \emph{replace}. 
In each update operation an XPath \emph{target} expression is used to specify the set of XML node(s) in which the update is applied. 
Moreover, a second argument \emph{source} is required for \emph{insert} and \emph{replace} operations which represents a sequence of XML nodes.
Note that \emph{target} may evaluate to an arbitrary sequence of nodes, denoted \emph{target}-\emph{nodes}, in case of \emph{delete} operation. 
As for other operations, however, \emph{target} must evaluate to a single node, denoted \emph{target}-\emph{node}; 
otherwise a dynamic error is raised. The XML update operations considered in this paper are detailed as follows:\medskip

\noindent $\bullet$ \textbf{insert} $source$ \textbf{into} / \textbf{as first into} / \textbf{as last into} / \textbf{before} / \textbf{after} $target$: 
Inserts each node in $source$ as child, as first child, as last child, as preceding sibling node, or as 
following sibling node of $target$-$node$ respectively. The order defined between nodes of $source$ must be preserved.
We abbreviate these kinds of $insert$ operations by \textit{\texttt{insertInto}}, \textit{\texttt{insertAsFirst}}, \textit{\texttt{insertAsLast}}, \textit{\texttt{insertBefore}}, 
and \textit{\texttt{insertAfter}} respectively. In case of \textit{\texttt{insertBefore}} and \textit{\texttt{insertAfter}} operations, 
$target$-$node$ must have a parent node; otherwise a dynamic error is raised. 
For \textit{\texttt{insertInto}} operation, the position of insertion is undetermined and may depend on the XUF implementation. 
Thus, the effect of executing an \textit{\texttt{insertInto}} operation on $target$ can be that of \textit{\texttt{insertAsFirst}}/\textit{\texttt{insertAsLast}} executed 
on $target$ or \textit{\texttt{insertBefore}}/\textit{\texttt{insertAfter}} executed at any child node of $target$.\medskip

\noindent $\bullet$ \textbf{delete} $target$: This operation is used to delete all nodes in \emph{target}-\emph{nodes} along with their 
descendant nodes.\medskip

\noindent $\bullet$ \textbf{replace} $target$ \textbf{with} $source$: Used to replace \emph{target}-\emph{node} and its descendants with the sequence of 
nodes specified in $source$ by preserving their order. Note that \emph{target}-\emph{node} must have a parent node; otherwise a dynamic error is raised.

\section{Update Access Control Model}\label{Sect3}
This section describes our access control model for XML update. 

\subsection{Update Specifications}
\label{updateSpecifications}
We focus on the security annotation principle presented in \cite{ref4} and on the \emph{update access type} notion introduced in \cite{ref20} 
to define our update specifications.

\begin{definition}\label{definition1}
Given a DTD $D$, an \emph{update type} defined over $D$ is of the form 
\texttt{insertInto}\emph{[$B_i$]}, \texttt{insertAsFirst}\emph{[$B_i$]}, \texttt{insertAsLast}\emph{[$B_i$]}, 
\texttt{insertBefore}\emph{[$B_i$]}, \texttt{insertAfter}\emph{[$B_i$]}, \texttt{delete}\emph{[$B_i$]} or \texttt{replace}\emph{[$B_i$,$B_j$]}, 
where $B_i$ and $B_j$ are element types of $D$.\hfill\(\square\)
\end{definition}

\noindent Intuitively, an update type $ut$ represents a set of update operations which are defined for specific element types. 
For example, the update type \textit{\texttt{insertInto}}[$B$] represents the update operations 
``\textbf{insert} $source$ \textbf{into} $target$'' where nodes in $source$ are of type $B$. Moreover, \textit{\texttt{replace}}[$B_i$,$B_j$] represents the update operations 
``\textbf{replace} $target$ \textbf{with} $source$'' where $target$-$node$ is of type $B_i$ and nodes in $source$ are of type $B_j$.

\begin{definition}\label{definition2}
We define an \emph{update specification} $S_{up}$ as a pair $(D,ann_{up})$, where $D$ is a DTD, and $ann_{up}$ is a partial mapping such that, 
for each element type $A$ in $D$ and each update type $ut$, $ann_{up}(A,ut)$, if defined, is an annotation of the form:
\begin{center}
 \texttt{$ann_{up}$($A$,$ut$) :=  $Y$ $|$ $N$ $|$ $[Q]$ $|$ $N_{h}$ $|$ $[Q]_{h}$}
\end{center}
\noindent with \texttt{$Q$} is a qualifier in our XPath fragment $\mathcal{X}$.\hfill\(\square\)
\end{definition}
An update specification $S_{up}$ is an extension of a document DTD $D$ associating update rights with element types of $D$.

Let $n$ be a node of type $A$ in an instantiation of $D$.
Intuitively, the authorization values $Y$, $N$, and \texttt{$[Q]$} indicate that, 
the user is \emph{authorized}, \emph{unauthorized}, or \emph{conditionally authorized} respectively to perform update operations of type 
$ut$ at $n$ (case of $insert$ operations) or over children nodes of $n$ (case of $delete$ and $replace$ operations).
For instance, the annotation $ann_{up}$($A$,\textit{\texttt{insertInto}}[$B$])=$Y$ specifies 
that the user can insert nodes of type $B$ as children nodes of $n$. However, the annotation $ann_{up}$($A$,\textit{\texttt{replace}}[$B_i$,$B_j$])=[$Q$] 
indicates that $B_i$ children of $n$ can be replaced by new nodes of type $B_j$ iff: $n\vDash Q$.
An annotation $ann_{up}$($A$,$ut$)=$value$ is said \emph{valid} at node $n$ iff: (\textit{i}) $value$=$Y$; or, 
(\textit{ii}) $value$=[$Q$]|[$Q$]$_h$ and $n\vDash Q$.

Our model supports \emph{inheritance} and \emph{overriding} of update privileges. 
If $ann_{up}$($A$,$ut$) is not explicitly defined, then an $A$ element \emph{inherits} the authorization of its parent node 
that concerns the same update type $ut$.
On the other hand, if $ann_{up}$($A$,$ut$) is explicitly defined it may \emph{override} the 
inherited authorization of $A$ that concerns the same update type $ut$. All update operations are not permitted by default.

\begin{example}\label{example1} 
Consider the following annotations defined over the hospital DTD (see the instance given in Fig. \ref{hospitalDATA}):\medskip

\begin{algorithmic}
\STATE $R_1$: $ann_{up}$($medicalFolder$,\textit{\texttt{delete}}[$treatment$])=$Y$
\STATE $R_2$: $ann_{up}$($analysis$,\textit{\texttt{delete}}[$treatment$])=$N$
\STATE $R_3$: $ann_{up}$($diagnosis$,\textit{\texttt{delete}}[$treatment$])=$Y$\medskip
\end{algorithmic}

\noindent $R_1$ indicates that the $treatment$ children of $medicalFolder$ nodes can be deleted (e.g., node $treatment_1$). 
$R_2$ overrides the delete authorization ($Y$) inherited from $medicalFolder$ and indicates that 
the $treatment$ children of $analysis$ nodes cannot be deleted, such as $treatment_2$ and $treatment_3$ nodes 
(node $treatment_3$ inherits the delete authorization ($N$) from its parent node $treatment_2$ since the annotation 
$ann_{up}$($treatment$,\texttt{\textit{delete}}[$treatment$]) is not explicitly defined). Similarly, $R_3$ overrides the delete authorization ($N$) of $analysis$ 
to allow deletion of the $treatment$ children of $diagnosis$ nodes (case of node $treatment_4$).\hfill\(\square\)
\end{example}

Finally, the semantic of the specification values $N_{h}$ and [$Q$]$_{h}$ is given as follows: 
The annotation $ann_{up}$($A$,$ut$)=$N_{h}$ indicates that, for a node $n$ of type $A$, update operations of type $ut$ cannot be 
performed at $n$ and no overriding of this authorization value is permitted for descendant nodes of $n$. For instance, if $n'$ is a descendant 
node of $n$ whose type is $A'$, then an update operation of type $ut$ cannot be performed at $n'$ even though $ann_{up}$($A'$,$ut$)=$Y$ is explicitly defined.
While, with the annotation $ann_{up}$($A$,$ut$)=[$Q$]$_{h}$, descendant nodes of an $A$ element can override this authorization value 
only if $Q$ is valid at this element. For instance, let $n$ and $n'$ be two nodes of type $A$ and $A'$ respectively, and consider the annotation $ann_{up}$($A'$,$ut$)=[$Q'$], then 
an update operation of type $ut$ can be performed at node $n'$ iff: $n' \vDash Q'$. Moreover, if the annotation 
$ann_{up}$($A$,$ut$)=[$Q$]$_{h}$ is defined and $n'$ is a descendant of $n$, then the annotation $ann_{up}$($A'$,$ut$)=[$Q'$] takes effect and an 
update operation of type $ut$ can be performed at node $n'$ iff: $n \vDash Q$ and $n' \vDash Q'$.
We call annotation with value $N_h$ or [$Q$]$_h$ as $downward$-$closed$ annotation.

\begin{example}\label{example2}
Suppose that the hospital wants to impose an update policy that authorizes the doctors to update (insertion, deletion,...) 
only data of patients having category '$A$', which are under department '$cardiology$' and not involved by clinical trial. 
We define formally this policy over an update type $ut$ as follows:\medskip

\begin{algorithmic}
\STATE $R_1$: $ann_{up}$($dept$,$ut$)=[$\downarrow$::$dname$/$text()$='$cardiology$']$_{h}$
\STATE $R_2$: $ann_{up}$($clinical$,$ut$)=$N_{h}$
\STATE $R_3$: $ann_{up}$($patient$,$ut$)=[$\downarrow$::$categ$/$text()$='$A$']\medskip
\end{algorithmic}

\noindent Consider the case where $ut$=\texttt{\textit{\texttt{insertInto}}}[$treatment$]. For a node $p$ of type $patient$, the annotation $R_3$ takes effect over data 
of $p$ only if $p$ is under cardiology department and outside of clinics 
($p$ has no ancestor node of type $clinical$); otherwise no insertion of $treatment$ nodes is permitted under node $p$ regardless its category. 
For the XML document presented in Fig. \ref{hospitalDATA}(a), insertions under nodes $patient_{3}$ and $patient_{4}$ are permitted 
(e.g., insert some $treatment$ nodes into $medicalFolder_3$).\hfill\(\square\)
\end{example}

In \cite{ref4}, when an annotation $R$ with qualifier [$Q$] is evaluated to false, all annotations under $R$ will be discarded regardless their truth values, 
which is not always necessary: 
According to our policy in Example \ref{example2}, insertions under node $patient_{3}$ of Fig. \ref{hospitalDATA}(a) are permitted by 
overriding negative authorization inherited from node $patient_{2}$. 
The principle of downward-closed annotation that we present with the specification values \{$N_h$,[$Q$]$_h$\} can be defined using just the values \{$Y$,$N$,$[Q]$\} but with 
a large XPath fragment (e.g., the fragment $\mathcal{X^{\Uparrow}}$) as done in \cite{ref6}. 
For instance, the policy of Example \ref{example2} can be defined as follows:\medskip

\begin{algorithmic}
\STATE $R'_1$: $ann_{up}$($dept$,$ut$)=[$\downarrow$::$dname$/$text()$='$cardiology$']
\STATE $R'_2$: $ann_{up}$($clinical$,$ut$)=$N$
\STATE $R'_3$: $ann_{up}$($patient$,$ut$)=[$\downarrow$::$categ$/$text()$='$A$' $and$ $not$($\uparrow^{+}$::$clinical$)\newline
\hspace*{11mm}$and$ $\uparrow^{+}$::$dept$[$\downarrow$::$dname$/$text()$='$cardiology$']]\medskip
\end{algorithmic}

\noindent Observe that, without using the values \{$N_h$,[$Q$]$_h$\}, defining a downward-closed annotation over element type $A$ amounts to propagate its value into all 
annotations defined under $A$ (values of the two downward-closed annotations $R_1$ and $R_2$ of Example \ref{example2} are propagated 
into the annotation $R_3$ to redefine it with $R'_3$). In this case, annotation $R'_3$ depends on $R'_1$ and $R'_2$ and must be redefined 
each time a modification is made on $R'_1$ and/or $R'_2$. This propagation leads to verbose annotations.
In other words, without the values \{$N_h$,[$Q$]$_h$\}, changing one annotation may require the modification of some annotations 
defined under it\footnote{This is not recommended for some systems like collaborative editing where the update policies are dynamic and 
each change is propagated to all the users across the network \cite{ImineCR09}.} which can be complicated and time consuming in case of large DTDs.

\subsection{DTD Recursion Problems}\label{dtdRecursionProblem}
Recall the problems 1 and 2 explained in Section \ref{Sect1}.
The first problem states the non-existence of models to specify update policies in case of recursive DTDs. 
Suppose that the hospital imposes that the doctors can 
update all treatment data except those which have been done outside the hospital (we suppose that all analysis are done in laboratories).
According to this policy, a doctor is permitted to update a $treatment$ node in an XML document 
(e.g., insert new $diagnosis$ data, delete some $result$ nodes, etc.) only if this node is not attached, directly or by other $treatment$ nodes, 
to an $analysis$ node. Given the XML document presented in Fig. \ref{hospitalDATA}, only nodes 
$treatment_1$ and $treatment_4$ can be updated while it is not the case of nodes $treatment_2$ and $treatment_3$ which are data analysis (attached to node $analysis_1$).
Such a policy can be defined only by using the notion of \emph{inheritance} and \emph{overriding} 
of update privileges which is not considered in the existing approaches \cite{ref3,ref14,ref19,ref20}.
This policy is defined in our model by the following update annotations:\medskip

\begin{algorithmic}
\STATE $R_1$: $ann_{up}$($medicalFolder$,$ut$)=$Y$
\STATE $R_2$: $ann_{up}$($diagnosis$,$ut$)=$Y$
\STATE $R_3$: $ann_{up}$($analysis$,$ut$)=$N$\medskip
\end{algorithmic}

\noindent where $ut$ can be any update type defined over element types $treatment$, $descp$, $diagnosis$, 
and $result$ (e.g., \textit{\texttt{delete}}[$result$], \textit{\texttt{insertInto}}[$diagnosis$]). 

The second problem is related to the enforcement of update policies. 
In case of recursive DTD, an update operation with $target$ defined in fragment $\mathcal{X}$ cannot be rewritten into an equivalent one 
defined in $\mathcal{X}$ in order to update only authorized data. This problem is known as the XPath closure problem \cite{ref5}. 
For instance, according to the previous update annotations, the update operation 
\textit{\texttt{delete}} $\downarrow^{+}$::$treatment$ cannot be rewritten into a safe update expressed in $\mathcal{X}$. Indeed, the paths denoting updatable treatment nodes 
(not done during analysis) stand for an infinite set. This set of paths can be captured with:
\textbf{delete} ($\downarrow^{+}$::$medicalFolder$ $\cup$ $\downarrow^{+}$::$diagnosis$)/($\downarrow$::$treatment$)*/$\downarrow$::$treatment$. 
However, the kleene star ($*$) cannot be expressed in XPath \cite{ref23,ref24}.

In the next section we explain how the extended fragment $\mathcal{X}_{[n]}^{\Uparrow}$, defined in Section \ref{Sect2}, can be used to overcome this 
\emph{update operations rewriting problem}.

\section{Secure Updating XML}\label{Sect4}
In this section we focus only on update rights and we assume that every node is read-accessible by all users.
Given an update specification $S_{up}$=$(D,ann_{up})$, we discuss the enforcement of such update constraints where 
each update operation posed over an instance $T$ of $D$ must be evaluated only over nodes of $T$ that can be updated by the user w.r.t $S_{up}$.
We assume that the XML document $T$ remains valid after the update operation is performed, otherwise the update is rejected.
In the following, we denote by $S_{ut}$ the set of annotations defined in $S_{up}$ over the update type $ut$ and by $|S_{ut}|$ 
the size of this set. Moreover, for a mapping function $ann$ (such as $ann_{up}$ of an update specification $S_{up}$=$(D,ann_{up})$), we denote by \{$ann$\} the set 
of all annotations defined with $ann$, and by |$ann$| the size of this set.

\subsection{Updatability}\label{UpdatabilitySect}
Consider the annotation $ann_{up}$($A$,$ut$)=$value$ and let $n$ be a node of type $A$. If this annotation is valid at $n$ 
then update operations of type $ut$ can change either the content of $n$ 
(i.e., delete/replace children nodes of $n$, insert new ones) or the 
information relative to its preceding-sibling (resp. following-sibling) presented by the relation $R_{\leftarrow}$ (resp. $R_{\rightarrow}$) in 
Section \ref{Sect2} (i.e., insert new nodes in preceding/following sibling of $n$).
Thus, we say that a node $n$ is \emph{updatable} w.r.t update type $ut$ if the user is granted to perform update operations of type $ut$ 
either at node $n$ (case of \emph{insert} operations) or over children nodes of $n$ (case of \emph{delete} and \emph{replace} operations).
For instance, if a node $n$ is updatable w.r.t \textit{\texttt{insertInto}}[$B$], then some nodes of type $B$ can be inserted as children of $n$. Additionally, 
$B_i$ children of $n$ can be replaced with nodes of type $B_j$ iff $n$ is updatable w.r.t \textit{\texttt{replace}}[$B_i$,$B_j$].

\begin{definition}\label{definition3}
Let $S_{up}$=$(D,ann_{up})$ be an update specification and $ut$ be an update type. 
A node $n$ in an instantiation of $D$ is \emph{updatable} w.r.t $ut$ if the following conditions hold:

\begin{itemize}
 \item[\textit{i})] The node $n$ is concerned by a valid annotation with type $ut$; or, 
no annotation of type $ut$ is defined over element type of $n$ and there is an ancestor node $n'$ of $n$ such that:
$n'$ is the first ancestor node of $n$ concerned by an annotation of type $ut$, and this annotation is valid at $n'$ (the inherited annotation).
 \item[\textit{ii})] There is no ancestor node of $n$ concerned by an invalid downward-closed annotation of type $ut$.\hfill\(\square\)
\end{itemize}
\end{definition}
 
\begin{example}\label{example3}
We consider the XML instance of Fig. \ref{hospitalDATA} and we define the following update annotations:\medskip

\begin{algorithmic}
\STATE $R_1$: $ann_{up}$($medicalFolder$,\textit{\texttt{insertInto}}[$result$])=$Y$
\STATE $R_2$: $ann_{up}$($diagnosis$,\textit{\texttt{insertInto}}[$result$])=$Y$
\STATE $R_3$: $ann_{up}$($analysis$,\textit{\texttt{insertInto}}[$result$])=$N$\medskip
\end{algorithmic}

\noindent The update \textbf{insert} <$result$/> \textbf{into} $\downarrow^{+}$::$treatment$[$\downarrow$::$descp$/$text()$='$biotherapy$'] 
has no effect since the node concerned by this update is $treatment_3$ which is not updatable w.r.t \textit{\texttt{insertInto}}[$result$]: 
According to Definition \ref{definition3}, no annotation of type \textit{\texttt{insertInto}}[$result$] is defined over 
element type $treatment$; and $analysis_1$ is the first ancestor node of $treatment_3$ concerned by an annotation of type 
\textit{\texttt{insertInto}}[$result$], annotation $R_3$. But, $R_3$ is not valid at $analysis_1$.\hfill\(\square\)
\end{example}

Given an update specification $S_{up}$=$(D,ann_{up})$, we define two predicates 
$\mathcal{U}^{1}_{ut}$ and $\mathcal{U}^{2}_{ut}$ (expressed in fragment $\mathcal{X}^{\Uparrow}_{[n]}$) to satisfy the conditions (\textit{i}) and (\textit{ii}) of 
Definition \ref{definition3} with respect to an update type $ut$:\medskip

\begin{algorithmic}
\STATE $\mathcal{U}^{1}_{ut}$ := $\uparrow^{*}$::$*$[$\bigvee_{(ann_{up}(A,ut)=Y|N|[Q]|N_h|[Q]_h)\in S_{ut}}$ $\varepsilon$::$A$][1]\newline
\hspace*{11mm}[$\bigvee_{(ann_{up}(A,ut)=Y)\in S_{ut}}$ $\varepsilon$::$A$ $\bigvee_{(ann_{up}(A,ut)=[Q]|[Q]_{h})\in S_{ut}}$ $\varepsilon$::$A[Q]$]\medskip

$\mathcal{U}^{2}_{ut}$ := $\bigwedge_{(ann_{up}(A,ut)=N_{h})\in S_{ut}}$ not ($\uparrow^{+}$::$A$)\newline 
\hspace*{11mm}$\bigwedge_{(ann_{up}(A,ut)=[Q]_{h})\in S_{ut}}$ not ($\uparrow^{+}$::$A[not (Q)]$)\medskip
\end{algorithmic}

\noindent where $\bigwedge$ and $\bigvee$ denote $conjunction$ and $disjunction$ respectively. 
The predicate $\mathcal{U}^{1}_{ut}$ has the form $\uparrow^{*}$::$*$[$qual_1$][1][$qual_2$]. 
Applying $\uparrow^{*}$::$*$[$qual_1$] on a node $n$ returns an ordered 
set $\mathcal{S}$ of nodes (node $n$ and/or some of its ancestor nodes) such that for each one an annotation of type $ut$ is defined over 
its element type. The predicate $\mathcal{S}$[1] returns either node $n$, if an annotation of type $ut$ is defined over its element type; 
or the first ancestor node of $n$ 
concerned by an annotation of type $ut$. 
Thus, to satisfy condition (\textit{i}) of Definition \ref{definition3}, it amounts to check that the node returned by $\mathcal{S}$[1] is 
concerned by a valid annotation of type $ut$, done by $\mathcal{S}$[1][$qual_2$] (i.e., $n\vDash \mathcal{U}^{1}_{ut}$). 
The second predicate is used to check that all downward-closed annotations of type $ut$ 
defined over ancestor nodes of $n$ are valid (i.e., $n\vDash \mathcal{U}^{2}_{ut}$).

\begin{definition}\label{definition4}
Let $S_{up}$=$(D,ann_{up})$, $ut$, and $T$ be an update specification, an update type and an instance of DTD $D$ respectively. 
We define the \emph{updatability predicate} $\mathcal{U}_{ut}$ which refers to an $\mathcal{X}^{\Uparrow}_{[n]}$ qualifier such that, a node $n$ on $T$ 
is \emph{updatable} w.r.t $ut$ iff $n\vDash\mathcal{U}_{ut}$, 
where $\mathcal{U}_{ut}$ $:=$ $\mathcal{U}_{ut}^{1} \bigwedge \mathcal{U}_{ut}^{2}$.\hfill\(\square\)
\end{definition}

For example, the XPath expression $\downarrow^{+}$::$*$[$\mathcal{U}_{ut}$] stands for all nodes which are updatable w.r.t $ut$. 
As a special case, if $S_{ut}=\phi$ then $\mathcal{U}_{ut}=false$.

\begin{example}\label{example4}
According to the update policy of Example \ref{example2}, the updatability predicate $\mathcal{U}_{ut}$ := $\mathcal{U}^{1}_{ut}$ $\bigwedge$ $\mathcal{U}^{2}_{ut}$ is 
defined with:\medskip

\begin{algorithmic}
\STATE $\mathcal{U}^{1}_{ut}$ := $\uparrow^{*}$::$*$[$\varepsilon$::$dept$ $\bigvee$ $\varepsilon$::$clinical$ $\bigvee$ $\varepsilon$::$patient$][1]\newline
\hspace*{11mm}[$\varepsilon$::$dept$[$\downarrow$::$dname$/$text()$='$cardiology$']\newline
\hspace*{11mm}$\bigvee$ $\varepsilon$::$patient$[$\downarrow$::$categ$/$text()$='$A$']]\medskip 

$\mathcal{U}^{2}_{ut}$ := not ($\uparrow^{+}$::$dept$[not ($\downarrow$::$dname$/$text()$='$cardiology$')])\newline
\hspace*{11mm}$\bigwedge$ not ($\uparrow^{+}$::$clinical$)\medskip
\end{algorithmic}
\sloppy{
\noindent Applying the predicate $\uparrow^{*}$::$*$[$\varepsilon$::$dept$ $\bigvee$ $\varepsilon$::$clinical$ $\bigvee$ $\varepsilon$::$patient$] over 
the node $medicalFolder_3$ of Fig. \ref{hospitalDATA}(a) returns the ordered set $\mathcal{S}$=\{$patient_3$, $patient_2$, $dept_1$\} of nodes 
(each one is concerned 
by an annotation of type $ut$); $\mathcal{S}$[1] returns $patient_3$; and the 
predicate [$\varepsilon$::$dept$[$\downarrow$::$dname$/$text()$='$cardiology$'] $\bigvee$ $\varepsilon$::$patient$[$\downarrow$::$categ$/$text()$='$A$']] 
is valid at $patient_3$. Thus $\mathcal{U}^{1}_{ut}$ is valid at node $medicalFolder_3$. 
Also, we can see that $medicalFolder_3 \vDash \mathcal{U}^{2}_{ut}$. Consequently, the node $medicalFolder_3$ is updatable w.r.t $ut$ 
(i.e., $medicalFolder_3 \vDash \mathcal{U}_{ut}$). 
This means that, in case of $ut$=\textit{\texttt{insertInto}}[$treatment$], the user is granted to insert nodes of type $treatment$ under node $medicalFolder_3$. 
However, if $ut$=\textit{\texttt{delete}}[$treatment$], then $treatment$ children of node $medicalFolder_3$ can be 
deleted (case of node $treatment_1$ of the instance of Fig. \ref{hospitalDATA}).\hfill\(\square\)
}
\end{example}

\begin{property}\label{property1}
For an update specification $S_{up}$=$(D,ann_{up})$ and an update type $ut$, the updatability predicate 
$\mathcal{U}_{ut}$ can be constructed in at most $O(|ann_{up}|)$ time.\hfill\(\square\)
\end{property}

\noindent \textsc{Proof.} Intuitively, for an update type $ut$, the definition of the set $S_{ut}$ depends on the parsing of all annotations 
of $S_{up}$ (i.e., the set \{$ann_{up}$\}) in $O(|ann_{up}|)$ time. The construction of each predicate $\mathcal{U}^{1}_{ut}$ and $\mathcal{U}^{2}_{ut}$ 
over annotations of $S_{ut}$ takes $O(|S_{ut}|)$ time. Thus, the predicate $\mathcal{U}_{ut}$ can be constructed in 
at most $O(|S_{ut}|+|ann_{up}|)$=$O(|ann_{up}|)$ time (since $|S_{ut}|\leq|ann_{up}|$).\hfill\(\square\)\medskip

\subsection{Update Operations Rewriting}\label{updateOperationsRewritingSection}
Finally, we detail here our approach for enforcing update policies based on the notion of \textit{query rewriting}. 
Given an update specification $S_{up}$=$(D,ann_{up})$. For any update operation with $target$ defined in the XPath fragment $\mathcal{X}$, we translate this 
operation into a safe one by rewriting its $target$ expression into another one $target'$ defined in the XPath fragment $\mathcal{X}^{\Uparrow}_{[n]}$, 
such that evaluating $target'$ over any instance of $D$ returns only nodes that can be updated by the user w.r.t $S_{up}$. 
We describe in the following the rewriting of each kind of update operation considered in this paper. 
We refer to DTD $D$ as a pair $(Ele,Rg,root)$, and to $source$ as a sequence of nodes of type $B$.\medskip

\noindent\textbf{Delete/Replace Operations.} According to our model of update, if the user holds the \textit{\texttt{delete}}[$A$] 
right on a node $n$ then he can delete children nodes of $n$ of type $A$.
Thus, given the update operation ``\textbf{delete} $target$'', for each node $n$ of type $A_i$ referred to by $target$, parent node $n'$ of $n$ must be updatable w.r.t \textit{\texttt{delete}}[$A_i$] 
(i.e., $n' \vDash \mathcal{U}_{delete[A_i]}$). To this end, the $target$ expression of \texttt{\textit{delete}} operations can be rewritten into: 
$target$[$\bigvee_{A_i\in Ele}$ $\varepsilon$::$A_i$[$\uparrow$::$*$[$\mathcal{U}_{delete[A_i]}$]]].

\noindent Consider now the update operation ``\textbf{replace} $target$ \textbf{with} $source$''. 
A node $n$ of type $A_i$ referred to by $target$ can be replaced with nodes in $source$ if its parent node $n'$ is updatable w.r.t \textit{\texttt{replace}}[$A_i$,$B$] 
(i.e., $n' \vDash \mathcal{U}_{replace[A_i,B]}$). 
Therefore, the $target$ expression of the replace operation can be rewritten into: 
$target$[$\bigvee_{A_i\in Ele}$ $\varepsilon$::$A_i$[$\uparrow$::$*$[$\mathcal{U}_{replace[A_i,B]}$]]].\medskip

\noindent\textbf{Insert as first into/as last into/before/after Operations.} Consider the update operation ``\textbf{insert} $target$ \textbf{as first into} $source$''.
For any node $n$ referred to by $target$, the user can insert nodes in $source$ at the first child position of $n$, regardless the type of $n$, 
provided that he holds the \textit{\texttt{insertAsFirst}}[$B$] right on this node (i.e., $n \vDash \mathcal{U}_{insertAsFirst[B]}$). 
To check this, the $target$ expression of the above update operation can be simply rewritten into:  
$target$[$\mathcal{U}_{insertAsFirst[B]}$]. The same principle is applied for the operations \textit{\texttt{insertAsLast}}, 
\textit{\texttt{insertBefore}}, and \textit{\texttt{insertAfter}}.\medskip

\noindent\textbf{Insert into Operation.} In the following we assume that: if a node $n$ is concerned by an annotation of type \textit{\texttt{insertInto}}[$B$], 
then this annotation implies \textit{\texttt{insertAsFirst}}[$B$] (resp. \textit{\texttt{insertAsLast}}[$B$]) rights 
for $n$, and \textit{\texttt{insertBefore}}[$B$] (resp. \textit{\texttt{insertAfter}}[$B$]) rights for children nodes of $n$ (inspired from \cite{ref3}). 
In other words, if one can(not) insert children nodes of types $B$ at any child position of some node $n$ as 
specified by some annotations of type \textit{\texttt{insertInto}}[$B$], then one can(not) insert nodes of type $B$ in the first and last child position of $n$ and 
in preceding and following sibling of children nodes of $n$ (unless if there is some annotations of type \textit{\texttt{insertAsFirst}}[$B$], \textit{\texttt{insertAsLast}}[$B$], \textit{\texttt{insertBefore}}[$B$], 
or \textit{\texttt{insertAfter}}[$B$] respectively that specify otherwise). 
Thus, one can execute the update operation ``\textbf{insert} $source$ \textbf{into} $target$'' over an XML tree $T$ iff: 
(\textit{i}) one has the right to execute update operations of type \textit{\texttt{insertInto}}[$B$] on the node $n$ ($n \in$ $T$\textlbrackdbl$target$\textrbrackdbl); 
and (\textit{ii}) no annotation \emph{explicitly prohibits} update operations of type \textit{\texttt{insertAsFirst}}[$B$]/\textit{\texttt{insertAsLast}}[$B$] on node $n$ 
(resp. \textit{\texttt{insertBefore}}[$B$]/\textit{\texttt{insertAfter}}[$B$] on children nodes of $n$). 
When condition (\textit{ii}) does not hold (e.g. update operations of type \texttt{\textit{insertAsFirst}} is explicitly denied), this leads to situation where 
there is a \emph{conflict} between \texttt{\textit{insertInto}} and other insert operations.\newline

The first condition is checked using the updatability predicate $\mathcal{U}_{insertInto[B]}$ (whether or not $n \vDash\mathcal{U}_{insertInto[B]}$). For the second condition, however, 
we define the predicate $\mathcal{U}^{-1}_{ut}$ over an update type $ut$ such that: for a node $n$, if $n \vDash\mathcal{U}^{-1}_{ut}$ then 
update operations of type $ut$ are \emph{explicitly forbidden} on node $n$. 
An update operation of type $ut$ is \emph{explicitly forbidden} at node $n$ iff at least one of the following conditions holds: 
\textit{a}) the node $n$ is concerned by an invalid annotation of type $ut$; \textit{b}) no annotation of type $ut$ is defined over element type of $n$ and there is an ancestor node $n'$ of $n$ such that:
$n'$ is the first ancestor node of $n$ concerned by an annotation of type $ut$, and this annotation is invalid at $n'$; 
\textit{c}) there is an ancestor node of $n$ concerned by an invalid downward-closed annotation of type $ut$.

More formally, for an update specification $S_{up}$=$(D,ann_{up})$, 
we define the predicate $\mathcal{U}^{-1}_{ut}$ := $Cnd_{a\vee b}$ $\bigvee$ $Cnd_{c}$ over 
an update type $ut$ with:\footnote{As a special case, if $S_{ut}=\phi$ then $\mathcal{U}^{-1}_{ut}=false$.}\medskip

\begin{algorithmic}
\STATE $Cnd_{a\vee b}$ := $\uparrow^{*}$::$*$[$\bigvee_{(ann_{up}(A,ut)=Y|N|[Q]|N_h|[Q]_h)\in S_{ut}}$ $\varepsilon$::$A$][1]\newline
\hspace*{5mm}[$\bigvee_{(ann_{up}(A,ut)=N|N_h)\in S_{ut}}$ $\varepsilon$::$A$ $\bigvee_{(ann_{up}(A,ut)=[Q]|[Q]_{h})\in S_{ut}}$ $\varepsilon$::$A[not (Q)]$]\medskip

$Cnd_{c}$ := $\bigvee_{(ann_{up}(A,ut)=N_{h})\in S_{ut}}$ $\uparrow^{+}$::$A$\newline
\hspace*{5mm}$\bigvee_{(ann_{up}(A,ut)=[Q]_{h})\in S_{ut}}$ $\uparrow^{+}$::$A[not (Q)]$\medskip
\end{algorithmic}


To resolve the conflict between \texttt{\textit{insertInto}} operation and other insert types, 
we define the predicate $CRP_B$ (``\emph{Conflict Resolution Predicate}'') over an element type $B$ as:\medskip

\begin{algorithmic}
\STATE $CRP_B$ := $\mathcal{U}^{-1}_{insertAsFirst[B]}$ $\bigvee$ $\mathcal{U}^{-1}_{insertAsLast[B]}$ $\bigvee$\newline
\hspace*{5mm}$\downarrow$::$*$[$\mathcal{U}^{-1}_{insertBefore[B]}$] $\bigvee$ $\downarrow$::$*$[$\mathcal{U}^{-1}_{insertAfter[B]}$]\medskip
\end{algorithmic}

\noindent For a node $n$, if $n\vDash CRP_B$ then at least the update operation \textit{\texttt{insertAsFirst}}[$B$] (resp. \textit{\texttt{insertAsLast}}[$B$]) is forbidden for node $n$ or 
\textit{\texttt{insertBefore}}[$B$] (resp. \textit{\texttt{insertAfter}}[$B$]) is forbidden for some children nodes of $n$. 
Finally, given the update operation ``\textbf{insert} $source$ \textbf{into} $target$'' over an XML tree $T$, one can insert nodes of type $B$ in $source$ to the node 
$n$ ($n \in$ $T$\textlbrackdbl$target$\textrbrackdbl) if and only if: $n \vDash \mathcal{U}_{insertInto[B]}\bigwedge not(CRP_B)$. 
Thus, the $target$ of the \textit{\texttt{insertInto}} operation can be rewritten into: $target$[$\mathcal{U}_{insertInto[B]} \bigwedge not (CRP_{B})$].\medskip

The overall complexity time of our rewriting approach of update operations can be stated as follows:

\begin{theorem}\label{theorem1}
For any update specification $S_{up}$=$(D,ann_{up})$ and any update operation $op$ (defined in $\mathcal{X}$), there exists 
an algorithm ``\texttt{\textit{Rewrite Updates}}'' that translates $op$ into a safe one $op'$ (defined in $\mathcal{X}_{[n]}^{\Uparrow}$) in at most $O(|ann_{up}|)$ time.\hfill\(\square\)
\end{theorem}

\begin{figure}[t!]
\textbf{Algorithm:} \textit{\texttt{Rewrite Updates}}\newline           
\begin{algo}
\begin{footnotesize}
\SetKwInOut{Input}{input}
\SetKwInOut{Output}{output}
\Input{An update specification $S_{up}$=$(D,ann_{up})$ and an update operation $op$.}
\Output{a rewritten of $op$ w.r.t $S_{up}$.\newline}

let $D$=$(Ele,Rg,root)$\;
let $op$ be defined with $target$ and optional sequence $source$ of nodes which conform to type $B$\;
\uCase{\emph{(}\textit{\texttt{delete}} operation\emph{)} \emph{\textbf{:}}}{
$target'$ := $target$[$\bigvee_{A_i\in Ele}$ $\varepsilon$::$A_i$[$\uparrow$::$*$[$\mathcal{U}_{delete[A_i]}$]]];
}\uCase{\emph{(}\textit{\texttt{replace}} operation\emph{)} \emph{\textbf{:}}}{
$target'$ := $target$[$\bigvee_{A_i\in Ele}$ $\varepsilon$::$A_i$[$\uparrow$::$*$[$\mathcal{U}_{replace[A_i,B]}$]]];
}\uCase{\emph{(}\textit{\texttt{insertAsFirst}} operation\emph{)} \emph{\textbf{:}}}{
$target'$ := $target$[$\mathcal{U}_{insertAsFirst[B]}$];\newline
//\textit{same principle for \textit{\texttt{insertAsLast}}, \textit{\texttt{insertBefore}}, and \textit{\texttt{insertAfter}} operations}\;
}\Case{\emph{(}\textit{\texttt{insertInto}} operation\emph{)} \emph{\textbf{:}}}{
$CRP_B$ := $\mathcal{U}^{-1}_{insertAsFirst[B]}$ $\bigvee$ $\mathcal{U}^{-1}_{insertAsLast[B]}$\newline 
$\bigvee$ $\downarrow$::$*$[$\mathcal{U}^{-1}_{insertBefore[B]}$] $\bigvee$ $\downarrow$::$*$[$\mathcal{U}^{-1}_{insertAfter[B]}$]\;
$target'$ := $target$[$\mathcal{U}_{insertInto[B]} \bigwedge not (CRP_{B})$];
}
replace $target$ of $op$ with $target'$\;
\Return $op$;
\end{footnotesize}
\end{algo}
\caption{XML Update Operations Rewriting Algorithm.}
\label{UpdatesRewrite}
\end{figure}

\noindent \textsc{Proof.} Our algorithm ``\texttt{\textit{Rewrite Updates}}'' for XML update operations rewriting is given in Fig. \ref{UpdatesRewrite}. 
As explained in Section \ref{updateOperationsRewritingSection}, for any update specification $S_{up}$=$(D,ann_{up})$ with DTD $D$=$(Ele,Rg,root)$, 
the securing of an update operation $op$ consists in the rewriting of its $target$ 
expression (defined in $\mathcal{X}$) into a safe one $target'$ (defined in $\mathcal{X}_{[n]}^{\Uparrow}$) in order to refer only to XML 
nodes that can be updated by the user w.r.t $S_{up}$. 
Proving that $target'$ can be defined in $O(|ann_{up}|)$ time is intuitive and based on the proof of Property \ref{property1}:\medskip

\noindent $\bullet$ A \texttt{\textit{delete}} operation can be rewritten by adding the following predicate 
[$\bigvee_{A_i\in Ele}$ $\varepsilon$::$A_i$[$\uparrow$::$*$[$\mathcal{U}_{delete[A_i]}$]]] to its $target$ expression. 
For each element type $A_i$ in DTD $D$, $S_{delete[A_i]}$ is a subset of \{$ann_{up}$\}, i.e., $\bigcup_{A_{i}\in Ele}$ $S_{delete[A_i]}$ $\subseteq$ \{$ann_{up}$\}. 
All these subsets can be computed by parsing only one time the set \{$ann_{up}$\}, i.e., in $O(|ann_{up}|)$ time.
Next, each sub-predicate $\mathcal{U}_{delete[A_i]}$ is defined over the subset $S_{delete[A_i]}$ in $O(|S_{delete[A_i]}|)$ time, 
and all sub-predicates used in line 4 of Fig. \ref{UpdatesRewrite} can be defined in $O(\sum_{i}|S_{delete[A_i]}|)$=$O(|ann_{up}|)$ time. 
Therefore, the predicate [$\bigvee_{A_i\in Ele}$ $\varepsilon$::$A_i$[$\uparrow$::$*$[$\mathcal{U}_{delete[A_i]}$]]] can be defined in at most $O(|ann_{up}|)$ time, 
which is the rewriting time of \texttt{\textit{delete}} operations. 
The same principle is applied for \texttt{\textit{replace}} operations.\medskip

\noindent $\bullet$ For an \texttt{\textit{insertAsFirst}} operation (resp. \texttt{\textit{insertAsLast}}, \texttt{\textit{insertBefore}}, and \texttt{\textit{insertAfter}}) defined with $source$ 
of nodes conform to type $B$, only one predicate is used to rewrite this operation; 
the predicate [$\mathcal{U}_{insertAsFirst[B]}$] is constructed in at most $O(|ann_{up}|)$ time.\medskip

\noindent $\bullet$ An \texttt{\textit{insertInto}} operation defined with $source$ of nodes conform to type $B$ is rewritten by adding the predicate 
[$\mathcal{U}_{insertInto[B]} \bigwedge not (CRP_{B})$] to its $target$ expression (line 11 of Fig. \ref{UpdatesRewrite}). The predicate 
$\mathcal{U}_{insertInto[B]}$ is constructed in at most $O(|ann_{up}|)$ time, while the predicate $CRP_B$ is based on the definition of some other predicates $\mathcal{U}^{-1}_{ut}$ for each update type $ut$ 
in \{\texttt{\textit{insertAsFirst\emph{[$B$]}}}, \texttt{\textit{insertAsLast\emph{[$B$]}}}, \texttt{\textit{insertBefore\emph{[$B$]}}}, 
\texttt{\textit{insertAfter\emph{[$B$]}}}\}. 
Similarly to the updatability predicate, the construction of each predicate $\mathcal{U}^{-1}_{ut}$ takes at most $O(|ann_{up}|)$ time (the same proof as Property \ref{property1}). 
Thus, the overall complexity time of the rewriting of 
\texttt{\textit{insertInto}} operations is $O(5*|ann_{up}|)$=$O(|ann_{up}|)$ time.\hfill\(\square\)

\section{Secure Updating XML over Security Views}\label{Sect5}
In the previous section we have supposed that all nodes are read-accessible which is not always the case. 
An XML document $T$ can be queried simultaneously by different users. 
For each class of users, some read constraints can be imposed 
to deny access to sensitive information on $T$. To enforce such constraints, most of existing works which deal with read-access control are based on 
the notion of \emph{Security Views}.
Abstractly, for each class of users, we annotate the used DTD $D$ with read-access constraints to specify accessibility conditions for nodes of instances of $D$. 
A security view is defined to be a pair $(D_v,\sigma)$ where: (\textit{i}) $D_v$ is the view of $D$ given to the users to represent the schema of all and only 
data they are able to see; and (\textit{ii}) $\sigma$ is a function, hidden from the users, and used to extract, for each instance $T$ of $D$, its \emph{virtual} view $T_v$ 
showing only accessible nodes. 
We investigate in this section the secure of update operations defined over (recursive) security views.

\subsection{Access Control for Recursive Views}\label{secureQueryingXML}
Given a security view $V$=$(D_v,\sigma)$, some works \cite{ref4,ref6,ref19} have proposed efficient algorithms to rewrite any user query formulated 
for $D_v$ to an equivalent one formulated for the original DTD $D$ to be finally evaluated over any instance of $D$. 
This query rewriting principle has to avoid the overhead of view materialization and maintenance.
However, only non-recursive views are considered (i.e., $D_v$ is non-recursive). Consider the XPath fragment $\mathcal{X}$ which is more used in practice, 
it has been shown in \cite{ref5} that query rewriting is not always possible under $\mathcal{X}$ in case of recursive security views. 

To overcome this limitation, we presented in \cite{ref18} a general approach to make XPath query rewriting possible under recursive security views.
We briefly discuss here the main principle of our approach.\medskip

Given a DTD $D$=$(Ele,Rg,root)$, we define for each class of users an \emph{access specification} $S$=$(D,ann)$ which specifies accessibility of XML nodes 
in intances of $D$. Formally, $ann$ is a partial mapping such that, for each production $A \rightarrow Rg(A)$ and each element type $B$ in $Rg(A)$, 
$ann(A,B)$, if explicitly defined, is an annotation of the form: \texttt{$ann$($A$,$B$) :=  $Y$|$N$|[$Q$]|$N_{h}$|[$Q$]$_{h}$} 
where [$Q$] is a qualifier in our XPath fragment $\mathcal{X}$.

The specification values $Y$, $N$, and [$Q$] indicate that the $B$ children of $A$ elements in an instance of $D$ 
are \emph{accessible}, \emph{inaccessible}, or \emph{conditionally accessible} respectively. 
If $ann(A,B)$ is not explicitly defined, then $B$ inherits the accessibility of $A$ (\emph{inheritance}). On the other hand, if $ann(A,B)$ is explicitly defined it may 
\emph{override} the accessibility inherited from $A$ (overriding). 

The same principle of downward-closed annotation defined in Section \ref{updateSpecifications} is applied for access annotations. 
With the annotation $ann(A,B)$=$N_h$, each $B$ child of an $A$ element is inaccessible and any descendant node of this $B$ element can override this 
accessibility value ($N_h$) to be accessible. However, with the annotation $ann$($A$,$B$)=[$Q$]$_{h}$, for any node $n$ of type $B$ child of an $A$ element, descendant nodes 
of $n$ can override this accessibility value ([$Q$]$_h$) only if $n\vDash Q$.

We define the security view in our approach to be $V$=$(D_v,ann)$ by omitting the function $\sigma$ 
since it cannot be defined in case of recursive DTDs as outlined in \cite{ref4,ref6}.\medskip

Finally, we describe our algorithm ``\texttt{\textit{Rewrite}}'' for XPath queries rewriting over arbitrary security views (recursive or not). 
Given an access specification $S$=$(D,ann)$, we extract first the security view $V$=$(D_{v},ann)$ corresponding to $S$. 
The user is provided with the DTD view $D_v$ which represents the schema of the data he is able to see. 
For any query $Q$ defined in $\mathcal{X}$ over $D_v$, our algorithm ``\texttt{\textit{Rewrite}}'' translates it into an equivalent one 
$Q_{t}$ defined in $\mathcal{X}^{\Uparrow}_{[n,=]}$ over the original DTD $D$ such that: for any instance $T$ of $D$, its virtual view $T_{v}$ conforms to $D_v$, 
the evaluation of $Q$ on $T_v$ yields the same result as the evaluation of $Q_t$ on $T$. 
Our rewriting algorithm ``\texttt{\textit{Rewrite}}'' runs in linear time on the size of the query.\medskip
 
We explain now some notations used in the following. For an access specification $S$=$(D,ann)$, we define predicates $\mathcal{A}^{acc}$ and $\mathcal{A}^{+}$ (expressed in fragment $\mathcal{X}^{\Uparrow}_{[n]}$) such that: 
for any node $n$ in an instance of $D$, $n$ is $accessible$ w.r.t $S$ if and only if $n\vDash\mathcal{A}^{acc}$. While, $n$/$\mathcal{A}^{+}$ returns all accessible ancestor nodes of $n$. 
We use algorithm ``\textit{\texttt{RW\_Pred}}'' to rewrite any predicate $p$ defined in $\mathcal{X}$ over DTD view $D_v$ to an equivalent one \textit{\texttt{RW\_Pred}}($p$) 
defined in $\mathcal{X}^{\Uparrow}_{[n,=]}$ over the original DTD $D$. 
More details of these predicates, and about the algorithms ``\textit{\texttt{Rewrite}}'' and ``\textit{\texttt{RW\_Pred}}'' can be found in \cite{ref18}.

\subsection{Securing Update Operations}
We present in this section a security view-based approach for securing XML update operations. 
Given an access specification $S$=$(D,ann)$, and its corresponding security view $V$=$(D_v,ann)$. 
The update privileges of each class of users are defined over the DTD view $D_v$=$(Ele_v,Rg_v,root)$ to be $S_{up}$=$(D_{v},ann_{up})$ and not over 
the original DTD $D$ (i.e., for an update type $ut$, an annotation $ann_{up}(A,ut)$=$value$ defined over an element type $A$ does not make sense if $A\notin Ele_v$).
Each update operation must be rewritten with respect to both $V$ and $S_{up}$ to be safe, since, 
considering only update privileges (i.e., rewriting update operations only over $S_{up}$ as explained in Section \ref{updateOperationsRewritingSection}) 
is not sufficient to make XML updates secure and can cause leakage of sensitive information hidden by $V$. 
We illustrate this problem by the following example.

\begin{figure}[t!]
\centering
\includegraphics[width=7.3cm]{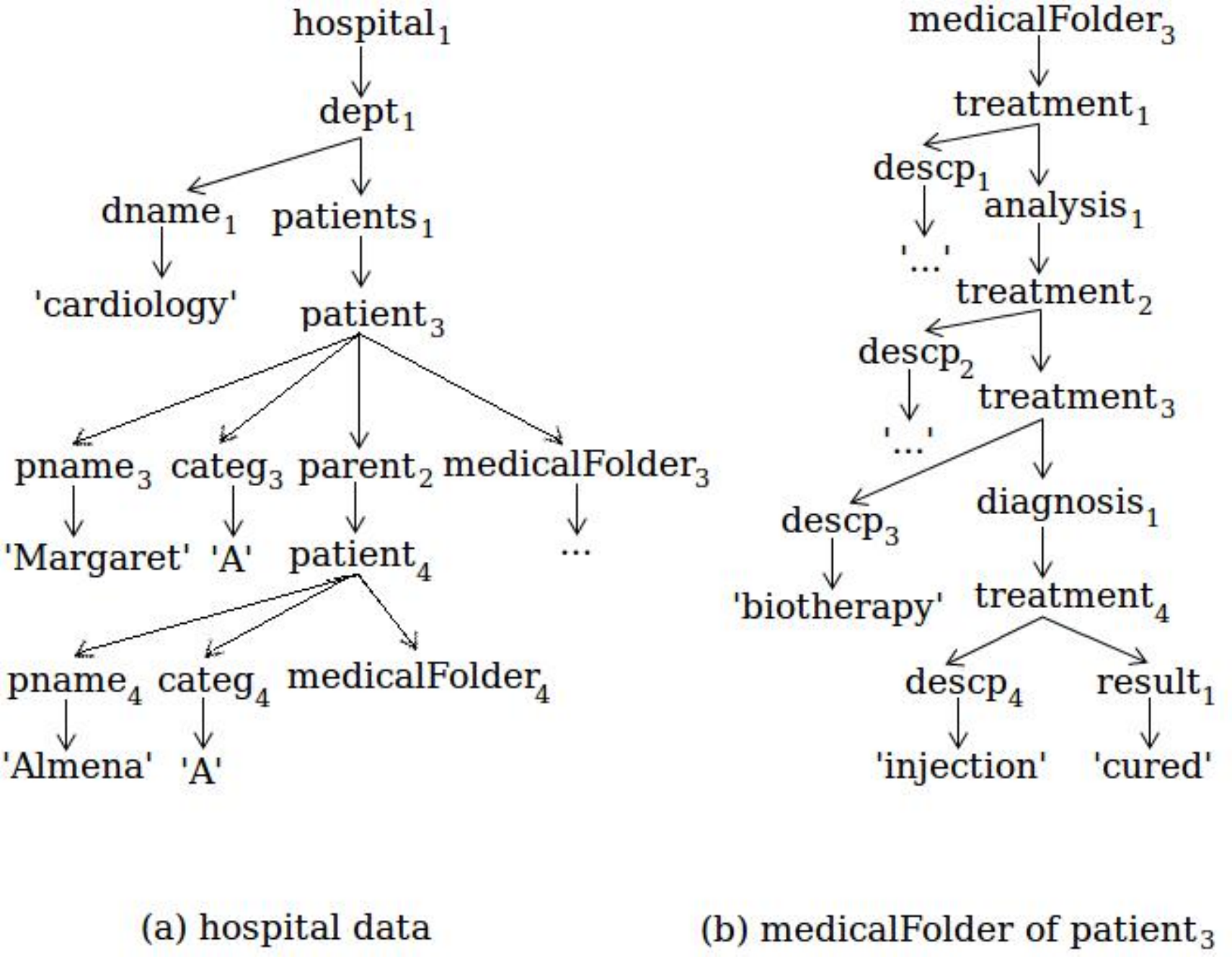}
\caption{View of the XML document of Fig. \ref{hospitalDATA}.}
\label{hospitalView}
\end{figure}

\begin{example}\label{example5}
Let $S$=$(D,ann)$ be an access specification where $D$ is the hospital DTD and the annotations $ann$ are defined as follows:

\begin{algorithmic}
\STATE $ann(hospital,dept)$=$[Q_{1}]_{h}$; $Q_{1}$ is $\downarrow$::$dname$/$text()$=$'cardiology'$\newline
$ann(dept,clinical)$=$N_{h}$
\STATE $ann(patients,patient)$=$[Q_2]$; $Q_{2}$ is $\downarrow$::$categ$/$text()$=$'A'$
\STATE $ann(parent,patient)$=$[Q_{2}]$
\end{algorithmic}

\noindent These annotations indicate that only the patients which are under department '$cardiology$', not involved by clinical trial, 
and also having category '$A$' are accessible to the user. Figure \ref{hospitalView} represents the virtual view of the XML instance of Fig. \ref{hospitalDATA} according to these annotations.
We define now the following update privileges:

\begin{algorithmic}
\STATE $ann_{up}(medicalFolder,\textit{\texttt{delete}}[result])=Y$
\STATE $ann_{up}(diagnosis,\textit{\texttt{delete}}[result])=Y$
\STATE $ann_{up}(analysis,\textit{\texttt{delete}}[result])=N$
\end{algorithmic}

\noindent \sloppy{Consider now the update operation $op$ = \textbf{delete} $\downarrow^{+}$::$patients$[$Q$]/$\downarrow^{+}$::$result$ defined over the view 
instance depicted in Fig. \ref{hospitalView} where $Q$ is the qualifier ``not ($\downarrow$::$patient$[$\downarrow$::$pname$/$text()$='$Margaret$'])''.
Considering only the update privileges is not sufficient to make this operation safe. 
The rewritten of this update operation w.r.t the update policy defined above returns $op'$=$op[\uparrow$::$*[\mathcal{U}_{delete[result]}]]$. 
If the execution of $op'$ over the original instance of Fig. \ref{hospitalDATA} deletes the node $result_1$, 
then the qualifier $Q$ is valid at node $patients_1$ and the user can deduce that some 
nodes are hidden between nodes $patients_1$ and $patient_3$.
By performing the rewritten operation $op'$ with $Q$=``$\downarrow$::$patient$[$\downarrow$::$pname$/$text()$='$Nathaniel$']'', 
the $result_1$ node is deleted and the user can deduce that patient $Nathaniel$ is currently residing in the hospital and has confidential data. 
Moreover, the user is able to request these sensitive data simply by changing the predicate $Q$.}\hfill\(\square\) 
\end{example}

In order to avoid this inference problem, each update operation must be rewritten w.r.t both read and update privileges to 
be safely executed over any instance. Securely controlling an update operation is then done in two steps:\medskip

\begin{algorithmic}
\STATE (1) The XPath $target$ expression of the update operation is rewritten according to the read privileges 
of the user submitting the update operation. This is done by using our rewriting algorithm ``\texttt{\textit{Rewrite}}'' described in Section \ref{secureQueryingXML}.\medskip

(2) Let $target'$ be the rewriting of $target$ w.r.t the read privileges, the user must hold the update privilege for each node referred to by $target'$. 
Then, we rewrite $target'$ w.r.t the update privileges into a safe one in order to be evaluated only over nodes updatable by the user and 
without disclosure of sensitive information.\medskip
\end{algorithmic}

\begin{example}\label{example6}
Consider the read and update privileges of Example \ref{example5}.
The update operation $op$=\textbf{delete} $\downarrow^{+}$::$patients$[$Q$]/$\downarrow^{+}$::$result$ 
(where $Q$ is the qualifier ``not ($\downarrow$::$patient$[$\downarrow$::$pname$/$text()$='$Margaret$'])'') over 
the view instance of Fig. \ref{hospitalView} is rewritten into \textbf{delete} $target''$ to be safely 
evaluated over the original instance of Fig. \ref{hospitalDATA}, where $target''$ is defined with:\medskip

\begin{algorithmic}
\STATE $target$ := $\downarrow^{+}$::$patients$[$Q$]/$\downarrow^{+}$::$result$\medskip

$target'$ := \texttt{\textit{Rewrite}}($target$) = \newline
\hspace*{5mm}$\downarrow^{*}$::$result$[$\mathcal{A}^{acc}$][$\uparrow^{+}$::$patients$[$\mathcal{A}^{acc}$][\texttt{\textit{RW\_Pred}}($Q$)][$\uparrow^{+}$::$hospital$]]\medskip

\texttt{\textit{RW\_Pred}}($Q$) := not ($\downarrow^{+}$::$patient$[$\mathcal{A}^{acc}$][$\downarrow^{+}$::$pname$[$\mathcal{A}^{acc}$]\newline
\hspace*{5mm}[$\varepsilon$::$*$/$text()$='$Margaret$']/$\mathcal{A}^{+}$[1]=$\varepsilon$::$patient$]/$\mathcal{A}^{+}$[1]=$\varepsilon$::$patients$)\medskip

$target''$ := $target'$[$\uparrow$::$*$[$\mathcal{U}_{delete[result]}$]]\medskip
\end{algorithmic}

\noindent We have seen in Example \ref{example5} that, by evaluating the predicate $Q$=``not ($\downarrow$::$patient$\newline[$\downarrow$::$pname$/$text()$='$Margaret$'])'' over node $patients_1$ of Fig. \ref{hospitalDATA}, 
some confidential information can be deduced. Using our rewriting algorithm ``\texttt{\textit{Rewrite}}'', we ensure that only accessible nodes can be requested by the update operation. 
Let $Q'$ be the predicate ``$\downarrow^{+}$::$patient$[$\mathcal{A}^{acc}$][$\downarrow^{+}$::$pname$[$\mathcal{A}^{acc}$][$\varepsilon$::$*$/$text()$='$Margaret$']\newline
/$\mathcal{A}^{+}$[1]=$\varepsilon$::$patient$]/$\mathcal{A}^{+}$[1]=$\varepsilon$::$patients$'' (i.e., \texttt{\textit{RW\_Pred}}($Q$)=not ($Q'$)).
Evaluating the predicate $Q'$ over a node $n$ in the original instance has to check that there is some accessible nodes of type $patient$, 
having name '$Margaret$', and which are children of $n$ or separated from it only with inaccessible nodes. 
Thus, the rewritten predicate \texttt{\textit{RW\_Pred}}($Q$) (i.e., not ($Q'$)) 
is not valid at node $patients_1$ since the node $patient_3$ has name '$Margaret$' and is separated from $patients_1$ only with inaccessible nodes. 
Therefore, the rewritten update operation \textbf{delete} $target''$ has no effect over the original instance of Fig. \ref{hospitalDATA} 
and no confidential information can be deduced.\hfill\(\square\)
\end{example}

\section{Conclusion}\label{Sect6}
We have proposed a general model for specifying XML update policies based on the primitives of XQuery Update Facility. 
To enforce such policies, we have introduced a rewriting approach to securely updating XML over arbitrary DTDs and for a significant fragment of XPath.
In the second part of this work, we have investigated the secure of XML data in the presence of security views. 
We have reviewed first our previously proposed approach enabling XPath query rewriting over recursive security views. 
Finally, we have discussed some inference problem that can be caused by combining read and update privileges, and our solution to deal with such a problem.
This yields the first XML security model that provides both read and update access control for arbitrary DTDs (resp. security views).

We plan first to extend our approach to handle larger fragments of XPath and other XQuery update operations. 
Moreover, we aim to provide a working system in order to investigate the practicality of our proposed solutions.

\scriptsize
\bibliographystyle{abbrv}
\bibliography{RR-7870.bib}

\begin{thebibliography}{10}

\bibitem{ref17}
A.~Berglund, S.~Boag, D.~Chamberlin, M.~F. Fern\'andez, M.~Kay, J.~Robie, and
  J.~Sim\'eon.
\newblock Xml path language (xpath) 2.0 (second edition).
\newblock {\em W3C Recommendation}, December 2010.

\bibitem{ref20}
L.~Bravo, J.~Cheney, and I.~Fundulaki.
\newblock Accon: checking consistency of xml write-access control policies.
\newblock In {\em EDBT}, pages 715--719, 2008.

\bibitem{ref19}
E.~Damiani, M.~Fansi, A.~Gabillon, and S.~Marrara.
\newblock A general approach to securely querying xml.
\newblock {\em Computer Standards {\&} Interfaces}, 30(6):379--389, 2008.

\bibitem{ref14}
M.~Duong and Y.~Zhang.
\newblock An integrated access control for securely querying and updating xml
  data.
\newblock In {\em ADC}, pages 75--83, 2008.

\bibitem{ref4}
W.~Fan, C.~Y. Chan, and M.~N. Garofalakis.
\newblock Secure xml querying with security views.
\newblock In {\em SIGMOD Conference}, pages 587--598, 2004.

\bibitem{ref13}
W.~Fan, F.~Geerts, X.~Jia, and A.~Kementsietsidis.
\newblock Smoqe: A system for providing secure access to xml.
\newblock In {\em VLDB}, pages 1227--1230, 2006.

\bibitem{ref5}
W.~Fan, F.~Geerts, X.~Jia, and A.~Kementsietsidis.
\newblock Rewriting regular xpath queries on xml views.
\newblock In {\em ICDE}, pages 666--675, 2007.

\bibitem{ref3}
I.~Fundulaki and S.~Maneth.
\newblock Formalizing xml access control for update operations.
\newblock In {\em SACMAT}, pages 169--174, 2007.

\bibitem{ref2}
I.~Fundulaki and M.~Marx.
\newblock Specifying access control policies for xml documents with xpath.
\newblock In {\em SACMAT}, pages 61--69, 2004.

\bibitem{ref9}
G.~Gottlob, C.~Koch, and R.~Pichler.
\newblock Efficient algorithms for processing xpath queries.
\newblock {\em ACM Trans. Database Syst.}, 30(2):444--491, 2005.

\bibitem{ref8}
B.~Groz, S.~Staworko, A.-C. Caron, Y.~Roos, and S.~Tison.
\newblock Xml security views revisited.
\newblock In {\em DBPL}, pages 52--67, 2009.

\bibitem{ref18}
M.~Houari and A.~Imine.
\newblock Secure querying of recursive xml views: A standard xpath-based
  technique.
\newblock {\em INRIA Research Report, NANCY, France}, Available at:
  \url{http://hal.inria.fr/hal-00646135/en}. December 2011.

\bibitem{ImineCR09}
A.~Imine, A.~Cherif, and M.~Rusinowitch.
\newblock A flexible access control model for distributed collaborative
  editors.
\newblock In {\em Secure Data Management}, 2009.

\bibitem{ref21}
F.~Jacquemard and M.~Rusinowitch.
\newblock Rewrite-based verification of xml updates.
\newblock In {\em PPDP}, pages 119--130, 2010.

\bibitem{ref27}
Y.~Koglin, G.~Mella, E.~Bertino, and E.~Ferrari.
\newblock An update protocol for xml documents in distributed and cooperative
  systems.
\newblock In {\em ICDCS}, pages 314--323, 2005.

\bibitem{ref26}
A.~Kundu and E.~Bertino.
\newblock A new model for secure dissemination of xml content.
\newblock {\em IEEE Transactions on Systems, Man, and Cybernetics, Part C},
  38(3):292--301, 2008.

\bibitem{ref6}
G.~M. Kuper, F.~Massacci, and N.~Rassadko.
\newblock Generalized xml security views.
\newblock In {\em SACMAT}, pages 77--84, 2005.

\bibitem{ref1}
M.~Murata, A.~Tozawa, M.~Kudo, and S.~Hada.
\newblock Xml access control using static analysis.
\newblock In {\em ACM Conference on Computer and Communications Security},
  pages 73--84, 2003.

\bibitem{NevenS06}
F.~Neven and T.~Schwentick.
\newblock On the complexity of xpath containment in the presence of
  disjunction, dtds, and variables.
\newblock {\em Logical Methods in Computer Science}, 2(3), 2006.

\bibitem{ref7}
N.~Rassadko.
\newblock Policy classes and query rewriting algorithm for xml security views.
\newblock In {\em DBSec}, pages 104--118, 2006.

\bibitem{ref11}
N.~Rassadko.
\newblock Query rewriting algorithm evaluation for xml security views.
\newblock In {\em Secure Data Management}, pages 64--80, 2007.

\bibitem{ref16}
J.~Robie, D.~Chamberlin, M.~Dyck, D.~Florescu, J.~Melton, and J.~Sim\'eon.
\newblock Xquery update facility 1.0.
\newblock {\em W3C Recommendation}, March 2011.

\bibitem{ref15}
P.~Samarati and S.~D.~C. di~Vimercati.
\newblock Access control: Policies, models, and mechanisms.
\newblock In {\em FOSAD}, pages 137--196, 2000.

\bibitem{ref12}
A.~Stoica and C.~Farkas.
\newblock Secure xml views.
\newblock In {\em DBSec}, pages 133--146, 2002.

\bibitem{ref24}
B.~ten Cate.
\newblock The expressivity of xpath with transitive closure.
\newblock In {\em PODS}, pages 328--337, 2006.

\bibitem{ref23}
B.~ten Cate and C.~Lutz.
\newblock The complexity of query containment in expressive fragments of xpath
  2.0.
\newblock {\em J. ACM}, 56(6), 2009.

\bibitem{ref10}
R.~Vercammen, J.~Hidders, and J.~Paredaens.
\newblock Query translation for xpath-based security views.
\newblock In {\em EDBT Workshops}, pages 250--263, 2006.

\end{thebibliography}
\nocite{*}

\end{document}